\documentclass[aps,pra,twocolumn,showpacs,groupedaddress,longbibliography]{revtex4}
\usepackage{graphicx}
\usepackage{hyperref}
\usepackage{bm,color}
\usepackage{dcolumn}
\usepackage{epsfig,amsbsy,bm,amssymb,amsmath}
\usepackage{tabu}
\usepackage{tabularx}
\renewcommand*{\[}{\begin{equation}}
\renewcommand*{\]}{\end{equation}}

\newcommand{\hs}{\hspace*}
\newcommand{\vs}{\vspace*}

\newcommand{\eref}[1] {(\ref{#1})}
\newcommand{\Eref}[1] {Eq.~(\ref{#1})}
\newcommand{\Fref}[1] {Fig. \ref{#1}}

\newcommand{\nn}{\nonumber}
\newcommand{\be}{\begin{equation}}
\newcommand{\ee}{\end{equation}}
\newcommand{\br}{\begin{eqnarray*}}
\newcommand{\er}{\end{eqnarray*}}
\newcommand{\ba}{\begin{eqnarray}}
\newcommand{\ea}{\end{eqnarray}}
\newcommand{\bp}{\begin{minipage}}
\newcommand{\ep}{\end{minipage}}
\newcommand{\bt}{\begin{tabular}}
\newcommand{\et}{\end{tabular}}

\newcommand{\bs}{\bigskip}
\newcommand{\ms}{\vs{-5mm}}
\newcommand{\mms}{\vs{-2.5mm}}

\newcommand{\isum}%
{\mathop{\hbox{$\displaystyle\sum\kern-13.2pt\int\kern1.5pt$}}}

\newcommand{\w}{\omega}

\begin{document}
\title{Strongly resonant RABBITT on lithium }

\date{\today}

\author{Anatoli~S.~Kheifets}
%\email{A.Kheifetts@anu.edu.au}
\bs

\affiliation{Research School of Physics, The
Australian National University, Canberra ACT 2600, Australia}

\begin{abstract}
The process of reconstruction of attosecond beating by interference of
two-photon transitions (RABBITT) can become resonant with a discrete
atomic level either in the intermediate or the final continuous
states. Experimental observations of the former [Phys. Rev. Lett. 104,
  103003 (2010)] and latter [Nat. Commun. 7, 10566 (2016)] resonant
processes revealed modification of only those parts of the
photoelectron spectrum that overlapped directly with the resonance. In
the lithium atom and other members of the alkali metal family, the
valence shell $ns\to np$ transition to the intermediate RABBITT state
affects the whole photoelectron spectrum in the final state. The
strong additional resonant channel modifies entirely the ionization
dynamics and opens direct access to the resonant phase of the
two-photon transitions which is common for various single and multiple
electron ionization processes.  Elucidation of this phase has wider
implications for strongly resonant laser-matter interaction.
\end{abstract}

\maketitle
%\end{document}

Valence shell dipole transitions are commonly used for optical
manipulation of alkali metal atoms. This includes optical pumping
\cite{Happer1972}, trapping \cite{Tannoudji1998}, and cooling
\cite{Phillips1998}.  These processes are of importance for many
quantum technologies such as metrology \cite{Derevianko2011},
information processing \cite{Reiserer2015}, computations
\cite{Saffman2010} and simulations \cite{Bloch2008}.  Lithium, the
lightest member of the alkali atom family, can be magneto-opically
trapped \cite{PhysRevA.83.023413}, cooled \cite{Sharma2018} and pumped
selectively to various $2p_m$ magnetic substates \cite{deSilva2020}.
These manipulations make lithium an ideal target for collision
\cite{Sharma2018} and strong laser physics 
\cite{PhysRevA.83.023413,deSilva2020} experiments.

The process of reconstruction of attosecond beating by interference of
two-photon transitions (RABBITT) \cite{MullerAPB2002,TomaJPB2002} has
become a widely used tool for attosecond chronoscopy of atoms
\cite{Pazourek2015}, molecules \cite{PhysRevLett.117.093001,Vos2018}
liquids \cite{Jordan2020} and solids \cite{Lucchini2015,Locher2015}.
In RABBITT, XUV driven primary ionization is augmented by secondary IR
photon absorption or emission. These two latter processes lead to the
same final continuous state whose population depends on the relative
phase of the absorption/emission amplitudes. Experimental access to
this phase makes it possible to obtain the timing information and to
resolve photoemission on the attosecond time scale. RABBITT can become
resonant with a discrete atomic level either in the intermediate or
the final continuous states. In the former process, a discrete atomic
state substitutes a missing continuous intermediate state that falls
below the ionization threshold. Such an under-threshold RABBITT (or
uRABBITT) has been observed in He \cite{SwobodaPRL2010} and Ne
\cite{PhysRevA.103.L011101}. Alternatively, the final continuous state
can be tuned to a Fano resonance. Such experiments were conducted on
He \cite{Gruson2016} and Ar \cite{Kotur2016,Cirelli2018}.  In both
cases, the resonance has a mild effect on the observed photoelectron
spectrum in the final state modifying only those parts that overlap
directly with the resonance.

In this Letter, we demonstrate a very strong modification of the whole
photoelectron spectrum in lithium when the $2s\to2p$ transition
becomes resonant with the intermediate RABBITT state. The strong
additional resonant channel modifies entirely the ionization dynamics
beyond its simple interpretation in terms of the relative
absorption/emission phase converted to the atomic time delay. 

\begin{figure}[t]
\hs{-0cm}
\epsfxsize=13cm
%\epsffile{EPS/Fig11.eps}
\epsffile{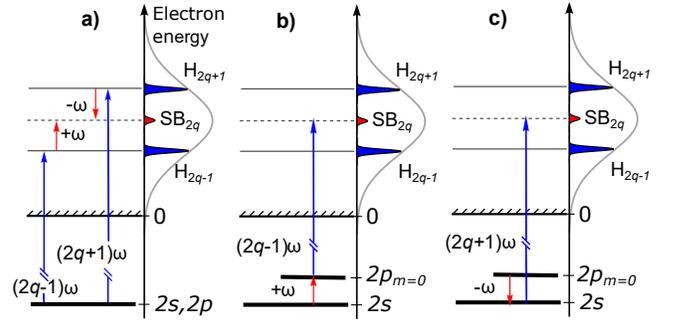}
\caption{(Color online) a) Schematic representation of the
  conventional RABBITT process on either of the Li $2s/2p$ initial
  state.  b) Same for the resonant Li $2s$ RABBITT process when the
  photon energy is in resonance with the level spacing $\w\simeq
  E_{2p}-E_{2s}$. c) Same for the resonant Li $2p_{m=0}$ process.
\label{Fig1}} 
\end{figure}

The conventional RABBITT on either of the Li $2s/2p$ initial states is
illustrated graphically in \Fref{Fig1}a). The atom absorbs an odd
number of the linearly polarized photons of the fundamental frequency
$(2q\pm1)\w$ to get ionized to one of the intermediate states which
are marked in the photoelectron spectrum by the harmonic order
H$_{2q\pm1}$. Subsequent emission or absorption of one IR photon leads
to the same final state which appears in the photoelectron spectrum as
a sideband SB$_{2q}$. The SB population oscillates when a time delay
$\tau$ is introduced between the  ionizing XUV and the dressing IR 
pulses
\ba
\label{Eq1}
S_{2q}(\tau) &=&
A+B\cos(2\omega\tau-C)
%\\
%C&=&
%\Delta\phi_{2q\pm1}+\Delta\phi_{\rm
%  W}+\Delta\phi_{\rm cc} 
%\ .
%\label{Eq2}
\ea
The simplest interpretation of the parameters entering \Eref{Eq1} is
provided by the lowest order perturbation theory (LOPT):
\ba
A&=&|{\cal M}_a|^2+|{\cal M}_e|^2
\  , \ 
B=2{\rm Re} \left[{\cal M}_a 
{\cal M}^*_e\right]
\nn\\
C&=& \arg\left[{\cal M}_{a}{\cal M}^*_e\right]
=2\w\tau_a \ .
\label{Eq2}
\ea
Here we introduce  the complex amplitudes of the XUV absorption, augmented
by absorption ${\cal M}_{a}$ or emission ${\cal M}_{e}$ of an IR photon. 
The phase of the RABBITT oscillation 
\be
\label{Eq3}
C=\Delta\phi_{2q\pm1}+\Delta\phi_{\rm
  W}+\Delta\phi_{\rm cc} 
\ee
is the sum of the phase difference of the neighbouring odd
harmonics ($\Delta\phi_{2q\pm1}=\phi_{2q+1}-\phi_{2q-1}$), the
analogous difference of the phases of the XUV absorption 
(the so-called Wigner phase difference) $\Delta\phi_{\rm W}$ and the
phase difference of the IR driven transitions (the so-called
continuum-continuum or CC phase difference) $\Delta\phi_{\rm cc}$. The
latter phase differences are converted to the corresponding time
delays by a finite difference formula \cite{Dahlstrom2012}
\be
\label{Eq4}
\tau_{\rm W}=\Delta\phi_{\rm W}/(2\w)
\  , \
\tau_{\rm cc}=\Delta\phi_{\rm cc}/(2\w)
\ .
\ee
The two time delays in \Eref{Eq4} add up to the atomic time delay
$\tau_a=\tau_{\rm W}+\tau_{\rm cc}$, which is the group delay of
the photoelectron wave packet propagating in the combined field of the
ion remainder and the dressing IR field relative to its free space
propagation.

The resonant $2s\to2p$ RABBITT process on the ground $2s$ state of Li
is illustrated in \Fref{Fig1}b). In this process, the IR photon
absorption promotes the electron to the $2p_{m=0}$ intermediate bound
state. Then the XUV $(2q-1)\w$ absorption elevates it to the same
final state SB$_{2q}$.  A similar resonant channel on the $2p_{m=0}$
initial state is exhibited in \Fref{Fig1}c. Here the IR photon is
first emitted and then the XUV $(2q+1)\w$ absorption populates
SB$_{2q}$. In both b) and c) the resonant RABBITT process does not
involve a CC transition and lacks the $\phi_{\rm cc}$ phase. Instead,
it contains the resonant phase which can be approximated by a
simplified expression \cite{PhysRevA.103.L011101} \mms
\be
\label{Eq5}
\phi_r\approx \arg \Big[ \w+E_{2s}-E_{2p}-i\Gamma\Big]^{-1}
=\arctan(\Gamma/\Delta)
\ .
\ee
Here $\Gamma$ is the spectral width of the IR pulse and $\Delta
\equiv\w+E_{2s}-E_{2p}$ is the detuning.  More elaborate expressions
for the resonant two-photon absorption phase are derived in
\cite{PhysRevA.93.023429,PhysRevLett.108.033003}.

\begin{figure}[t]
\vs{-0.5cm}
\epsfxsize=8.5cm
%\epsffile{delay/Phases/Li/ECROSS.eps}
\epsffile{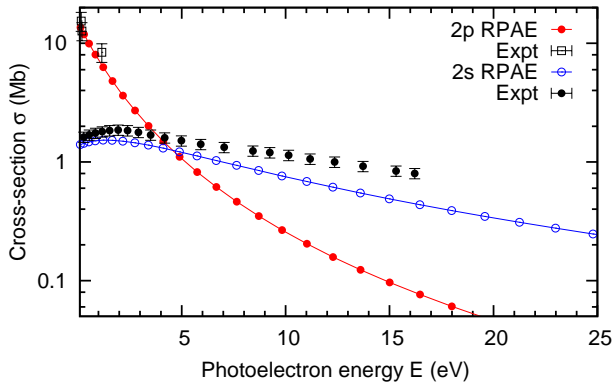}
\caption{(Color online) Partial photoionization cross-sections of the
  Li atom. Present calculations within the random phase approximation
  with exchange (RPAE) \cite{A90} are compared with the experimental
  data for the Li atom in the $2s$ \cite{Hudson1967} and $2p$
  \cite{Amin2006,Saini2019} initial states.
\label{Fig2}} 
\end{figure}

The relative contribution of the resonant and non-resonant RABBITT
processes and their phases depends on the strength of the
corresponding ionization channels. This strength can be gauged from
the partial photoionization cross-sections exhibited in \Fref{Fig2}.
In the limits of the small and large photoelectron energy $E$, these
cross-sections satisfy the following relations:
\begin{align}
\label{CS}
\sigma_{2p}&=14~{\rm   Mb}\gg\sigma_{2s}=1.3~{\rm Mb} & \rm for~&
E\simeq 0  
\\ \nn
\sigma_{2p}&\propto E^{-9/2}\ll\sigma_{2s}\propto E^{-7/2} &\rm for~&
E\gg I_p
\end{align}
The above relations show that the $2p$ primary ionization is much
stronger than the $2s$ one when the photoelectron energy is low.  When
this energy is high, it is the $2s$ primary ionization  that is
dominant over the $2p$ one. 
The $2s$ and $2p$ resonant channels exhibited in \Fref{Fig1}b) and c)
are driven by the $(2q-1)\w$ and $(2q+1)\w$ XUV photon absorption,
respectively. The former process approaches the threshold closely for
the lower SB orders whereas the latter process always stays away from
the threshold. Accordingly, the resonant channel of \Fref{Fig1}b)
weakens away from the threshold relative to the non-resonant $2s$
channel exhibited in \Fref{Fig1}a). Conversely, the resonant process
exhibited in \Fref{Fig1}c) is uniformly strong for all the SB's in
comparison with its non-resonant $2p$ counterpart.  Therefore the
resonant phase in this channel is dominant over the non-resonant one.
Notably, the $2p_{m=1}$ initial state does not mix with the
intermediate $2s_{m=0}$ state and the corresponding RABBITT process
lacks the resonant phase in this case.

\begin{figure}[t]

\hs{-7cm}{\bf a)} 

\vs{-0.5cm}\hs{1cm}
\epsfxsize=\columnwidth
%\epsffile{2P/N=11/1.55eV/3D1.eps}
\epsffile{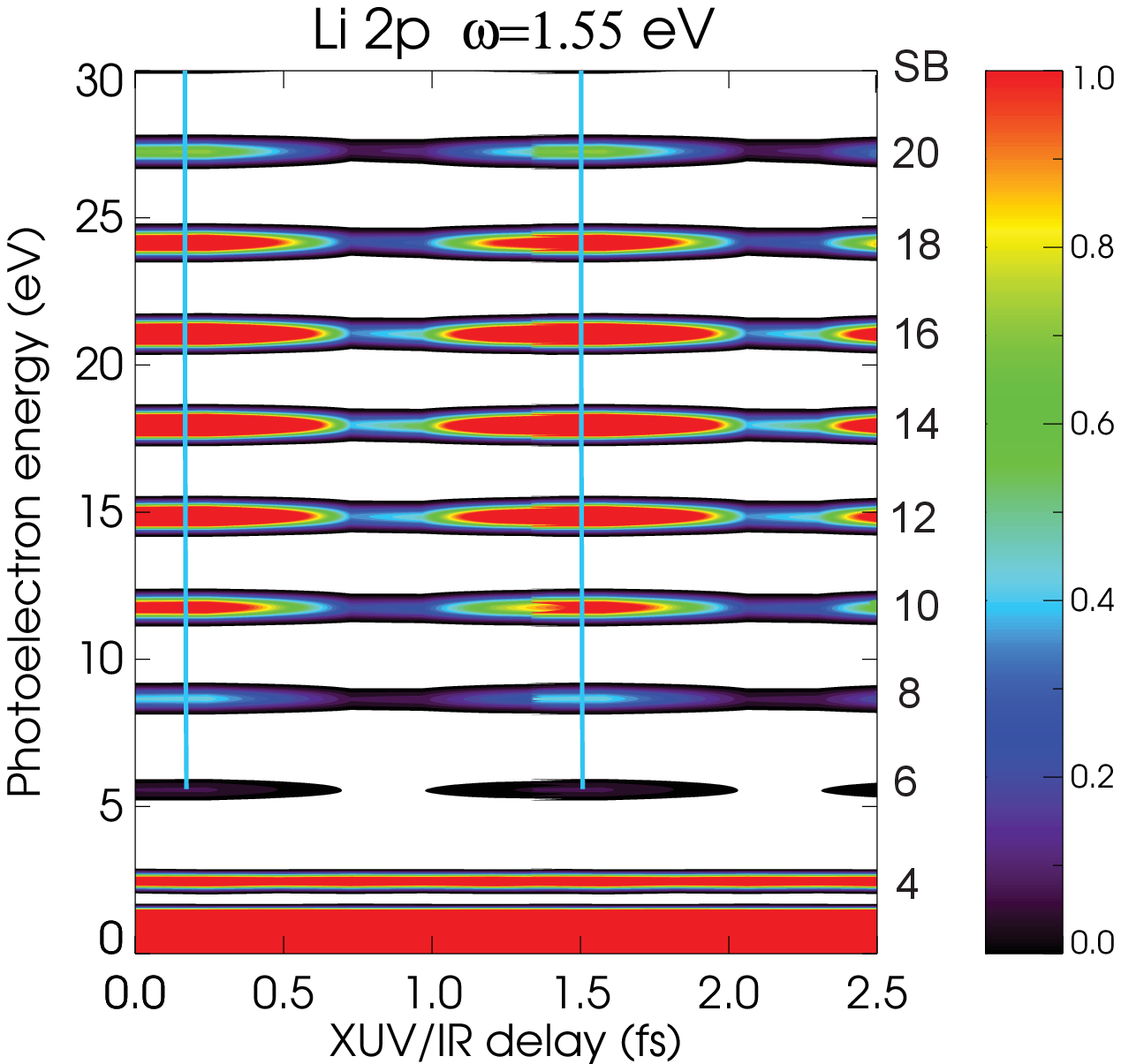}

\hs{-7cm}{\bf b)} 

\vs{-0.5cm}\hs{1cm}
\epsfxsize=\columnwidth
%\epsffile{2S/N=11/1.55eV/3D1.eps}
\epsffile{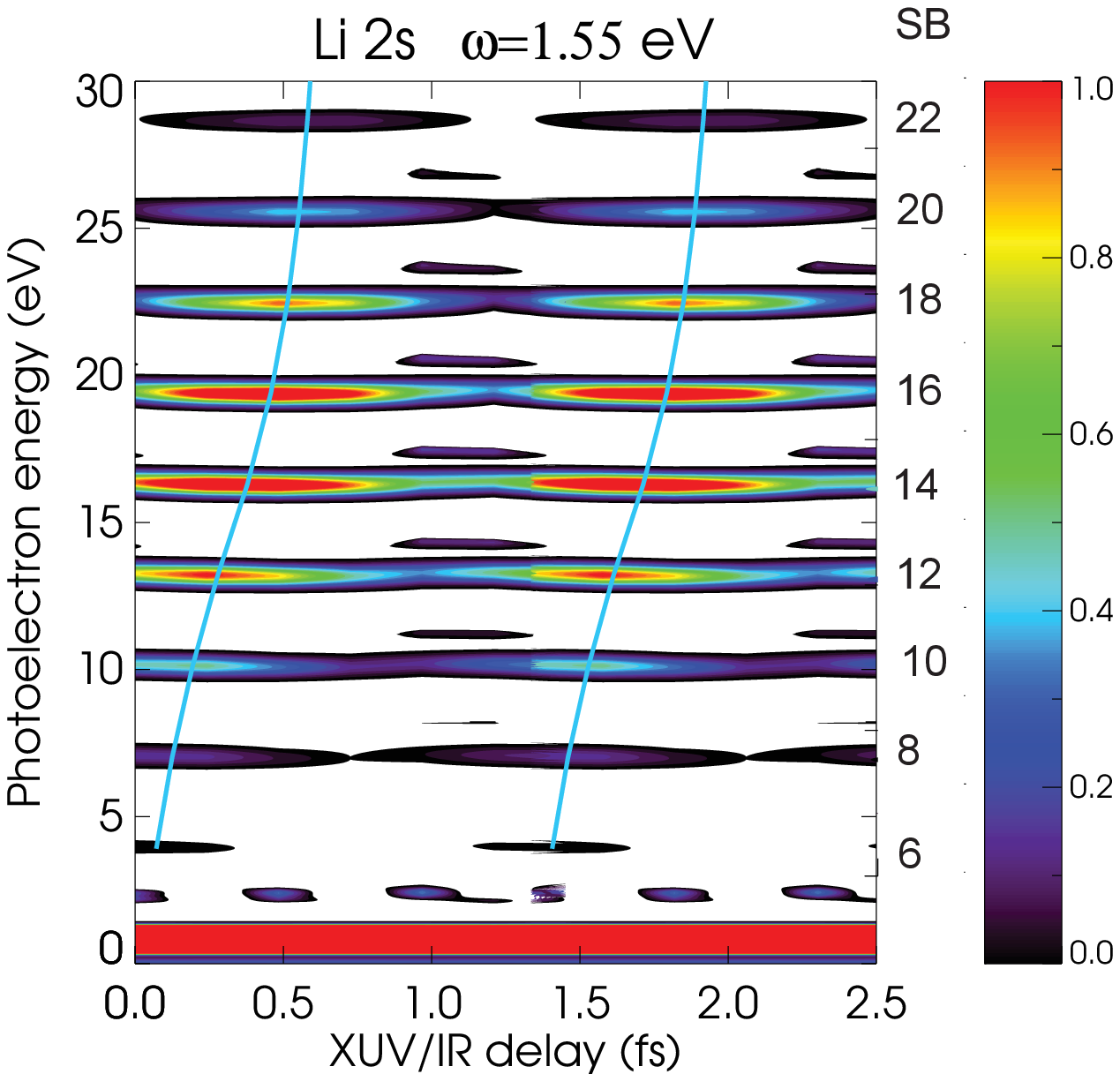}

\hs{-7cm}{\bf c)} 

\vs{-0.5cm}\hs{1cm}
\epsfxsize=\columnwidth
%\epsffile{2S/N=11/1.65eV/3D1.eps}
\epsffile{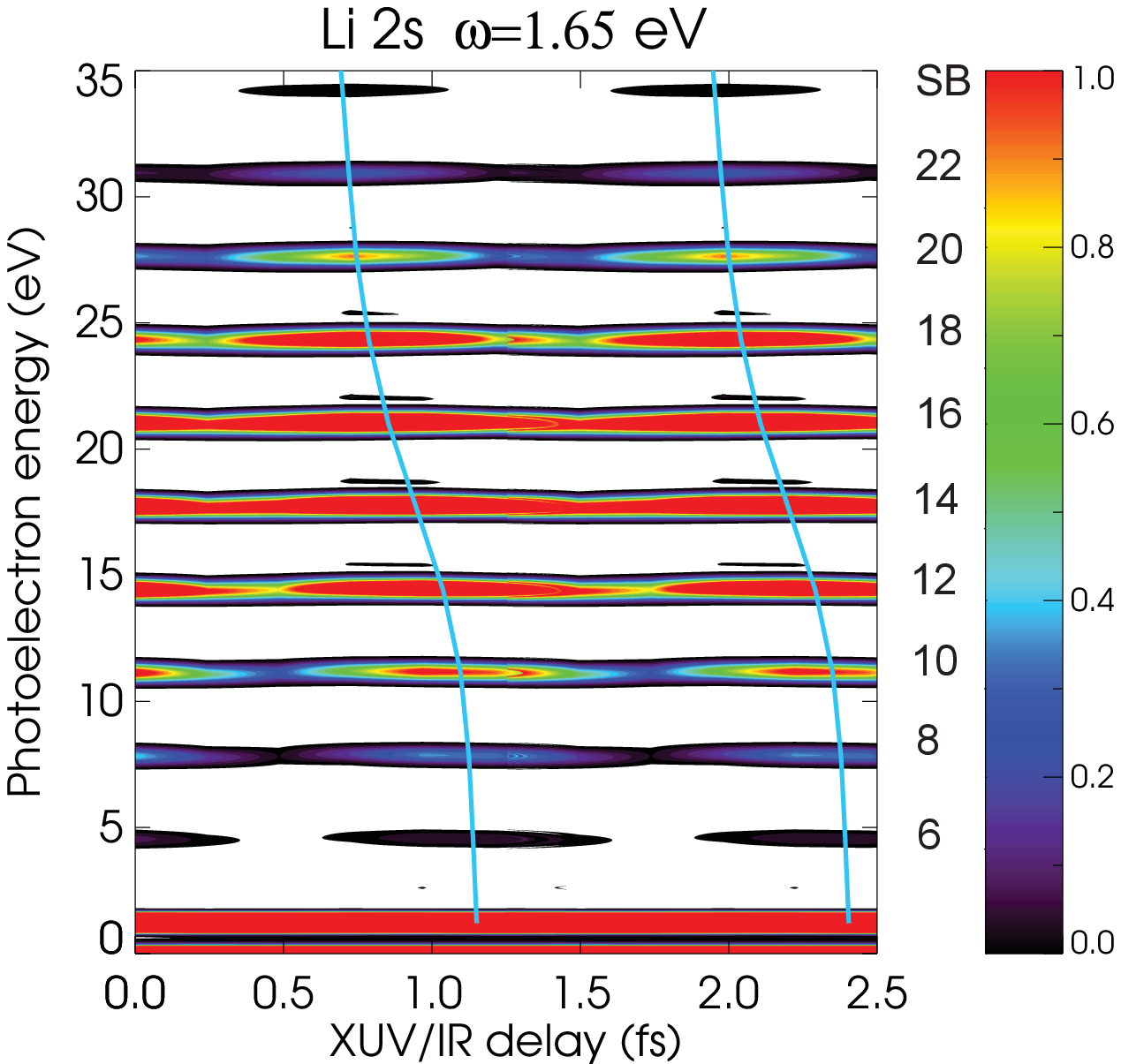}

\caption{(Color online) RABBITT traces of the Li atom initially in the
  $2p$ (a) and $2s$ (b,c) states at the fundamental frequency
  $\w=1.55$~eV, $\lambda=800$~nm (a,b) and $\w=1.65$~eV,
  $\lambda=729$~nm (c).  The corresponding SB orders are marked on the
  right vertical axis of each panel. The blue solid lines guide the
  eye through the centers of the sidebands.
\label{Fig3}} 
\end{figure}

Accurate non-perturbative treatment of the RABBITT process requires
numerical solution of the time-dependent Schr\"odinger equation
(TDSE). We seek this solution in the single-active electron (SAE)
approximation \cite{0953-4075-49-24-245001} with an effective
one-electron potential \cite{Sarsa2004163}. This approximation is
valid in the photon energy considered here which is well below the
$1s$ threshold at $\simeq 60$~eV. The TDSE SAE approach to RABBITT has
been tested successfully on He \cite{PhysRevA.94.063409}, Ne
\cite{PhysRevA.96.013408} and heavier noble gas atoms
\cite{PhysRevA.97.063404}.  The TDSE is driven by a superposition of
an XUV attosecond pulse train (APT) and the IR pulse in several fixed
increments of the IR/XUV delay $\tau$. 

The APT is modeled with the vector potential
\ba
\label{vectorGauss}
A_x(t) &=& \sum_{n=-5}^5 (-1)^n A_n \exp\left(
-2\ln2
{(t-nT/2)^2\over \tau_x^2}
\right)\nonumber\\
&&\times
\cos\Big[\omega_x(t-nT/2)\Big]  \ ,
\ea
where
$$
A_n = A_0
\exp\left(-2\ln2
{(nT/2)^2\over \tau_T^2}\right) \ .
$$
Here $A_0$ is the vector potential peak value
%
%related to the field intensity $ I = \omega^2|A_0|^2/$
%
and $T=2\pi/\omega$ is the period of the IR field.
The XUV central frequency is $\omega_x$ and the 
time constants $\tau_x, \tau_T$ are chosen  to span a sufficient
number of harmonics in the range of photon frequencies of 
interest for a given atom.

The vector potential of the IR pulse is represented by the cosine
squared envelope
\be
\label{vectorSin2}
A(t) = A_0 \cos^2
\left(
%{\pi (t-\Delta)\over 2\sqrt2\tau}
{\pi (t-\tau)\over 2\tau_{\rm IR}}
\right)
\cos[\omega(t-\tau)] \ .
\ee

In the present work, the APT is centered at $\w_x=15\w$ and its
spectral width $\Gamma=0.4$~eV.  Typical XUV and IR field intensities
are $5\times10^{9}$ and $3\times10^{10}$ W/cm$^2$ respectively. In
this low intensities regime, our numerical results depend weakly on
variation of these parameters.

The photoelectron
spectrum is obtained by projecting the time-dependent wave function at
the end of the time evolution on the basis of Volkov states. Numerical
details are given in the preceding
publications~\cite{PhysRevA.96.013408,PhysRevA.97.063404}.

Results of our simulations are shown in \Fref{Fig3} for the Li atom
initially in the $2p_m$ state summed over $m=0,\pm1$ (a) and the $2s$
state (b,c). The photon energy $\w=1.55$~eV in (a,b) and 1.65~eV in
(c). In \Fref{Fig3} we display the RABBITT traces which are comprised
of the stack of angular integrated photoelectron spectra taken while
varying the XUV/IR delay $\tau$. As the two-photon RABBITT transitions
are weaker than the one-photon primary photoionization, the harmonic
peaks are normally much stronger than the sidebands (see \Fref{Fig1}
for graphical illustration). To highlight the SB's, we conduct yet
another computation with the XUV ionization only and subtract the
resulting photoelectron spectrum from the XUV/IR RABBITT spectra at
each time delay.  Thus the primary harmonic peaks are all but removed
and the RABBITT traces of \Fref{Fig3} display the sidebands very
clearly.

The sidebands are integrated over the energy window $2q\w\pm \Gamma/2$
and their time dependence is fitted with \Eref{Eq1}. The resulting
phases $C_{2q}$ for each SB are marked on the RABBITT traces and
joined by the solid blue lines. These lines guide the eye through the
SB centers on each panel of \Fref{Fig3}. The striking difference
between the panels (a) and (b,c)  is that the SB's are
perfectly aligned in the case of the $2p$ initial state whereas they
are visibly tilted for the $2s$ initial state. The direction of this
tilt is opposite for the photon energies of 1.55~eV and 1.65~eV.

The lack of a SB dispersion for the $2p$ initial state can be
understood from \Eref{Eq3}. In our simulations, the APT is composed of
the pulselets of altering polarity and $\Delta\phi_{2q\pm1} =
\pi~\forall q$. In comparison, both the Wigner $\Delta\phi_{\rm W}$
and the CC $\Delta\phi_{cc}$ phase differences are small away from the
threshold. Also the resonant phase $\phi_r$ \eref{Eq5} does not depend
on the photoelectron energy $E$.  Thus the resulting RABBITT phase is
nearly constant for all the SB's.

\begin{figure}[t]
\hs{-7.5cm}{\bf a)} 

\vs{-0.5cm}
\epsfxsize=8.5cm
%\epsffile{2P/N=11/PLOT0.eps}
\epsffile{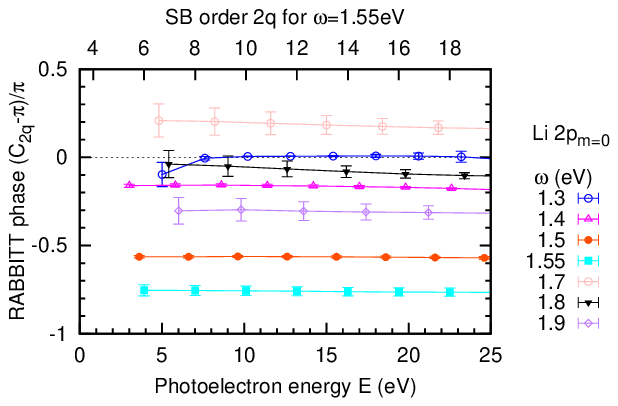}

\hs{-7.5cm}{\bf b)} 

\vs{-0.15cm}
\epsfxsize=8.5cm
%\epsffile{2P/N=11/PLOT1.eps}
\epsffile{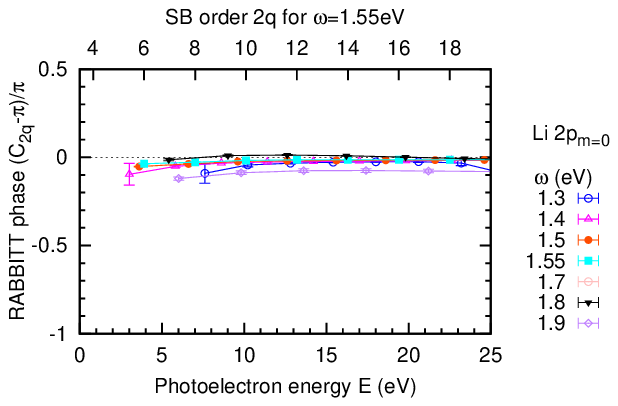}

\hs{-7.5cm}{\bf c)} 

\vs{-0.15cm}
\epsfxsize=8.5cm
%\epsffile{2S/N=11/PLOT1.eps}
\epsffile{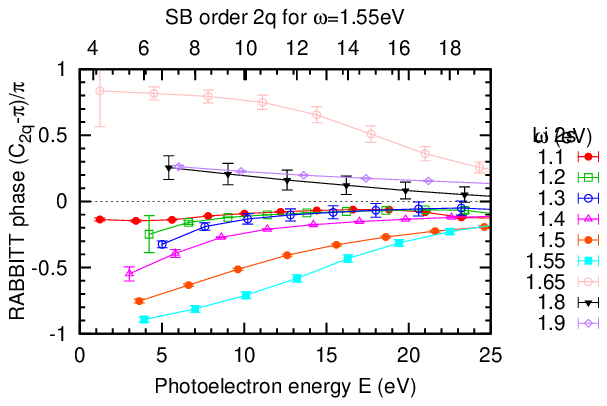}

\caption{(Color online) a) The $2p_{m=0}$ RABBITT phases $C_{2q}$ for
  various sidebands are plotted for several fixed photon energies. The
  top scale indicates the SB order for $\w=1.55$~eV. b) and c) is the
  same for the $2p_{m=1}$ and $2s$ RABBITT phases, respectively. 
\label{Fig4}} 
\end{figure}

The phase variation with the photoelectron energy and the fundamental
photon frequency $\w$ is analyzed in more detail in \Fref{Fig4}. In
the three panels of this figure, from top to bottom, we display the
net RABBITT phase corresponding to the $2p_{m=0}$, $2p_{m=1}$ and $2s$
initial states, respectively.  The harmonic phase difference
$\Delta\phi_{2q\pm1} = \pi$ is subtracted for clarity.  The
fundamental photon frequency $\omega$ varies across the resonant
$2s-2p$ transition. For the $2p_{m=0}$ initial state, the net RABBITT
phase depends strongly on $\w$ but remains flat with $E$. Conversely,
the $2p_{m=1}$ phase does not depend neither on $\w$ nor on $E$. And,
finally, the $2s$ phase depends strongly both on $\w$ and $E$. The
sign of the $E$ dispersion depends on $\w$. It turns from positive to
negative when $\w\ge1.65$~eV. This transition corresponds to the
photon energy approaching the level spacing $E_{2p}-E_{2s}$ (1.68~eV
in our model potential and 1.86~eV in the experiment \cite{NIST-ASD}).

\begin{figure}[t]

\epsfxsize=8.5cm
%\epsffile{2P/N=11/PLOT8.eps}
\epsffile{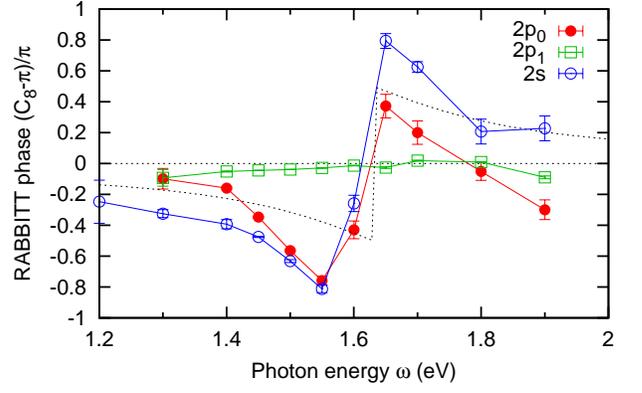}

\caption{
The $C_{8}$
  phase variation with the photon energy $\w$ is plotted for
  $2p_{m=0,1}$ and $2s$ initial states. The dotted line
  visualizes \Eref{Eq5}.
\label{Fig5}} 
\end{figure}

The resonant transition of the $2p_{m=0}$ and $2s$ RABBIT phases is
shown more distinctively in \Fref{Fig5} where we select just a single
SB$_8$ and trace its phase as a function of the photon energy. This
low SB$_8$ is dominated by the resonant channel for both the initial
states. For the $2p_{m=0}$ initial state, this resonant character is
retained by the higher order SB's and their phases remain nearly flat
over an extended range of the photoelectron energy $E$ as displayed in
\Fref{Fig4}a). Conversely, for the $2s$ initial state, the resonant
character of the higher order SB's weakens and their phases 
approach $\Delta\phi_{2q\pm1} = \pi$ as exhibited in
\Fref{Fig4}c). Thus the corresponding RABBITT phases demonstrate a
significant energy dispersion with $E$. 

\begin{figure}
\hs{-7.5cm}{\bf a)}

\vs{-0.4cm}
\epsfxsize=8cm
%\epsffile{2P/N=11/1.55eV/SB/CvalsM0/PLOT.eps}
\epsffile{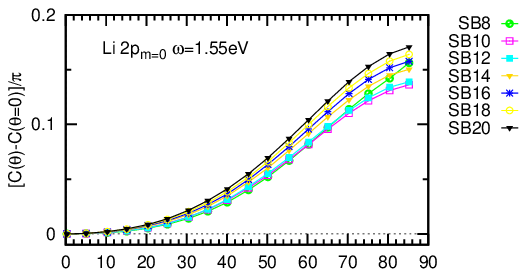}

\ms
\hs{-7.5cm}{\bf b)}

\vs{-0.25cm}
\epsfxsize=8cm
%\epsffile{2P/N=11/1.55eV/SB/CvalsM1/PLOT.eps}
\epsffile{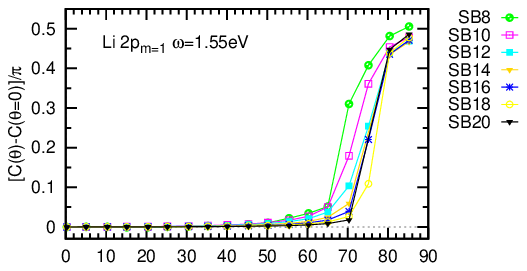}

\ms
\hs{-7.5cm}{\bf c)}

\vs{-0.25cm}
\epsfxsize=8cm
%\epsffile{2S/N=21/1.55eV/SB/Cvals/PLOT.eps}
\epsffile{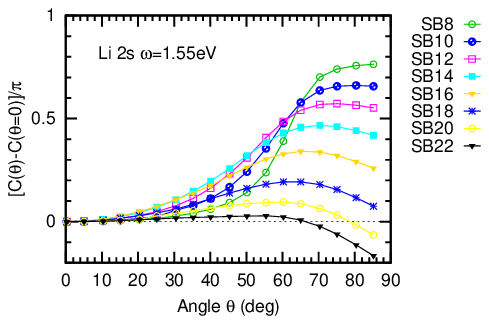}

\caption{(Color online) Variation of the RABBITT phase relative to
  the polarization direction $C(\theta)-C(\theta=0)$  on the Li $2p_{m=0}$ (a),
  $2p_{m=1}$ (b) and $2s$ (c) initial states. The photon energy $\w=1.55$~eV.
\label{Fig6}} 
\ms
\end{figure}

Up to now, we examined the angular integrated RABBITT spectra. The
angular dependence of these spectra can also be explored by tracing
the SB position and deducing its phase as a function of the
photoelectron emission angle $\theta$. This tracing is exhibited in
\Fref{Fig6} for the $2p_{m=0}$ (a), $2p_{m=1}$ (b) and $2s$ (c) initial
states. Here the RABBITT phase $C(\theta)-C(\theta=0)$ is plotted
relative to the polarization direction corresponding to $\theta=0$. In
the case of the $2p_{m=1}$ initial state, which is not resonant with
its  $2s$ counterpart, there is no angular variation of the
RABBITT phase except its sharp rise above $\theta\simeq60^\circ$. 
The smooth and rather uniform angular dependence appears in the
RABBITT phase for the $2p_{m=0}$ initial state which is consistently
resonant with the $2s$ state for all the SB's. 
In both cases, the angular dependence
originates from the competition of the two continuous final states,
$$
2p\stackrel{(2q\pm1)\w}{\longrightarrow} \epsilon d \stackrel{\mp \w}{\longrightarrow} Ep,Ef
\ ,
$$
each supported by their own spherical harmonics. The population of the
$\epsilon s$ intermediate state is 10 times smaller and can be
neglected. The $Ef$ channel should normally dominate over the $Ep$ one
because of the Fano propensity rule \cite{PhysRevA.32.617}, which was
confirmed in other RABBITT studies
\cite{PhysRevLett.123.133201,PhysRevA.103.L011101}. A single dominant
spherical harmonic does not provide any angular dependence below its
kinematic node.  This is indeed the case for the non-resonant
$2p_{m=1}$ initial state where the angular dependence is missing for
all the SB's below the node of $Y_{31}(63.4^\circ)$.  For the resonant
$2p_{m=0}$ initial state, the angular dependence is also uniform but
it is noticeable for all the emission angles. This is the evidence of
several competing spherical harmonics.  Finally, for the $2s$ initial
state, the angular dependence is strong for lower order SB's but it
gradually weakens for higher order SB's. For a non-resonant $ns$
initial state, it is the competition of the $\epsilon p\stackrel{ \pm
  \w}{\longrightarrow} Es,Ed$ transitions that introduces the angular
dependence of the RABBITT phase \cite{PhysRevA.94.063409}. Such a
dependence, however, reveals itself only beyond the kinematic node
$Y_{20}(54.7^\circ)$. In the present case of the resonant $2s$ initial
state, the angular dependence onsets at significantly smaller
angles. The angular dispersion is strong and positive for small SB's.
It becomes weak and negative for higher SB orders which is a typical
for a non-resonant He $1s$ initial state \cite{PhysRevA.94.063409}.

In conclusion, we studied systematically the RABBITT processes in the
Li atom prepared initially in the $2s$, $2p_{m=0}$ and $2p_{m=1}$
states. The properties of the RABBITT process in lithium are very
diverse. Experimentally, the population of the Li atom in various
$2p_m$ sub-states is achieved by resonant pumping by the linearly
\cite{PhysRevA.83.023413} or circularly \cite{deSilva2020} polarized
laser pulses.  The three initial states demonstrate different
interplay between the resonant and non-resonant RABBITT processes. The
contribution of the resonant channel is selective for the $2s$ initial
state. It is very strong for the lower order SB's near the threshold
but it weakens as the SB order and the corresponding photoelectron
energy grow. Such a variable competition between the resonant and
non-resonant channels leads to a strong SB energy and angular
dispersion.  The sign of the energy dispersion changes abruptly when
the photon energy passes through the resonance corresponding to the
energy spacing between the $2s$ and $2p$ initial states.  In the case
of the $2p_{m=0}$ initial state, the resonant channel makes a
uniformly dominant contribution for all the SB orders. In this case,
the energy dispersion of the SB's is very weak while the angular
dependence is uniform and moderate. Finally, the resonant channel does
not contribute for the $2p_{m=1}$ initial state. In this case, both
the energy and angular dispersion of the SB's are absent. It needs to
be stressed that a strong resonant channel invalidates the
conventional definition of the atomic time delay via the RABBITT phase
by way of \Eref{Eq2}. Indeed, the resonant phase is contained only in
one of the $(2q\pm1)\w$ XUV absorption arms and the finite difference
expression \eref{Eq4} for the phase derivative cannot be used.

Our study broadens significantly the catalogue of the resonant RABBITT
processes reported so far in the literature. In the previous studies,
only one selected sideband was affected by the resonance either in
the intermediate state (the so-called uRABBITT process
\cite{SwobodaPRL2010,PhysRevA.103.L011101}) or by tuning it to a Fano
resonance in the final state \cite{Gruson2016,Kotur2016,Cirelli2018}.

The strongly resonant RABBITT should be found in other members of the
alkali atoms family. Their valence shell dipole $ns\to np$ transitions
overlap with NIR laser frequencies and make these atoms convenient
targets for optical manipulation.  Importantly, because of the
identical principle quantum numbers, the oscillator strength of the
$ns\to np$ transition is several times greater than that of the higher
order transitions. This makes the resonant behaviour of the $ns$
RABBITT very robust and clear. In the meantime, a non-resonant $np_{m=1}$
RABBITT can serve as a stable reference which displays no sideband
dispersion except the high order harmonics group delay
(the attochirp). The latter instrumental effect is identical for both
initial states and thus can be easily eliminated.

Significance of the present findings goes beyond the specificity of
the RABBITT process. It opens direct access to the resonant phase of
the two-photon transitions. The resonant phase can be extracted
straightforwardly by taking the difference between the $np_{m=0}$ and
$np_{m=1}$ RABBITT measurements. This phase is common for various
single and multiple electron ionization processes. Several theoretical
models describing this phase
\cite{PhysRevA.93.023429,PhysRevLett.108.033003,PhysRevA.103.L011101}
can thus be validated. This will have wider implications for strongly
resonant laser-matter interaction.  The combination of the RABBITT and
magneto-optical trapping is technically challenging at
present \footnote{Thomas Pfeifer, Daniel Fischer, private
  communications (2021).}.  Nevertheless, we hope that we provided
sufficient motivation for such an experiment to be conducted. The
alkali metal atoms are the natural candidates for future attosecond
studies once the traditional noble gas targets are exhausted.

The author thanks Alex Bray for automating repetitive RABBITT
calculations with the use of efficient shell scripts.  Resources of
the National Computational Infrastructure (NCI) facility were utilized
in the present work.

%\np%~~\np
%\bibliography{references,ureferences,mreferences,sreferences,treferences,dreferences,mypapers,reft}

\end{document}